\documentclass[a4paper]{article}

\usepackage{INTERSPEECH2022}
\usepackage{acronym}
\usepackage[binary-units=true]{siunitx}
\usepackage{comment}
\usepackage[acronym,smallcaps]{glossaries}

\usepackage[english]{babel}
\usepackage[T1]{fontenc}
\usepackage[utf8]{inputenc}
\usepackage{booktabs}

\usepackage{xargs}    
\usepackage[pdftex,dvipsnames]{xcolor}  
\usepackage[colorinlistoftodos,prependcaption,textsize=tiny]{todonotes}

\newcounter{todocounter}
\newcommandx{\change}[2][1=inline]{\stepcounter{todocounter}\todo[linecolor=red,backgroundcolor=red!25,bordercolor=red,#1]{\thetodocounter: #2}}
\newcommandx{\info}[2][1=inline]{\stepcounter{todocounter}\todo[linecolor=OliveGreen,backgroundcolor=OliveGreen!25,bordercolor=OliveGreen,#1]{\thetodocounter: #2}}
\newcommandx{\improve}[2][1=inline]{\stepcounter{todocounter}\todo[linecolor=blue!25,backgroundcolor=blue!25,bordercolor=blue,#1]{\thetodocounter: #2}}
\newcommandx{\unsure}[2][1=inline]{\stepcounter{todocounter}\todo[linecolor=yellow,backgroundcolor=yellow!25,bordercolor=yellow,#1]{\thetodocounter: #2}}

\newcommand{\db}[1]{$\SI{#1}{dB}$}

\newcommand{\kbps}[1]{$\SI{#1}{kbps}$}
\newcommand{\khz}[1]{$\SI{#1}{kHz}$}
\newcommand{\ms}[1]{$\SI[per-mode=symbol,per-symbol = /]{#1}{\milli\second}$}
\newcommand{\seconds}[1]{$\SI[per-mode=symbol,per-symbol = /]{#1}{\second}$}

\acrodef{ourmodel}[NESC]{Neural End-2-End Speech Codec}

\acrodef{frontend}[DPCRNN]{DualPathConvRNN}
\acrodef{ssmgan}[SSMGAN]{Streamwise-StyleMelGAN}
\acrodef{vqvae}[VQ-VAE]{Vector-Quantized VAE}

\acrodef{evs}[EVS]{EVS}
\acused{evs}
\acrodef{lpcnet}[LPCNet]{LPCNet}
\acused{lpcnet}
\acrodef{lyra}[Lyra]{Lyra}
\acused{lyra}
\acrodef{opus}[OPUS]{OPUS}
\acused{opus}
\acrodef{sstream}[SoundStream]{SoundStream}
\acused{sstream}

\title{NESC: Robust Neural End-2-End Speech Coding with GANs}
\name{Nicola Pia, Kishan Gupta, Srikanth Korse, Markus Multrus, Guillaume Fuchs}
\address{Fraunhofer IIS Erlangen}
\email{\{nicola.pia, kishan.gupta, srikanth.korse, markus.multrus, guillaume.fuchs\}@iis.fraunhofer.de}

\begin{document}

\maketitle

\begin{abstract}
Neural networks have proven to be a formidable tool to tackle the problem of speech coding at very low bit rates.
However, the design of a neural coder that can be operated robustly under real-world conditions remains a major challenge.
Therefore, we present \ac{ourmodel} a robust, scalable end-to-end neural speech codec for high-quality wideband speech coding at \kbps{3}.
The encoder uses a new architecture configuration, which relies on our proposed \ac{frontend} layer, while the decoder architecture is based on our previous work \acl{ssmgan}.
Our subjective listening tests on clean and noisy speech show that \ac{ourmodel} is particularly robust to unseen conditions and signal perturbations.

\end{abstract}


\section{Introduction}
\label{sec:introduction}

Very low bit rate speech coding is extremely challenging for classical coding techniques.
The paradigm usually employed is parametric coding, which yields intelligible speech at the cost of poor audio quality and unnatural synthesized speech.
Recent advances in neural networks are filling this gap, enabling high-quality speech coding at very low bit rates.
We categorize the possible solutions to this problem according to the role played by neural networks.

\begin{itemize}
\item[level 1] \emph{post-filtering}: A neural network based post-processor is employed at the end of a conventional encoder-decoder chain, in order to improve the quality of the coded speech. This enables the enhancement of existing communication systems with minimal effort.
\item[level 2] \emph{neural decoder}: A conventional encoder model generates a bitstream, which is decoded using a neural network. This enables backward compatible decoding of existing bitstreams.
\item[level 3] \emph{end-2-end}: Both encoder and decoder are neural networks, which are trained jointly. The input of the encoder is the speech waveform, and the quantization is jointly learned, hence obtaining directly the optimal bitstream for the signal.
\end{itemize}

Level 1 approaches such as~\cite{cnn-postfilter, opus-postfilter, mask-postfilter, gan-postfilter, postgan} are minimally invasive, as they can be deployed over existing pipelines.
Unfortunately they still suffer typical unpleasant artifacts, which are especially challenging to eliminate.

The first published level 2 speech decoder was based on WaveNet~\cite{wavenet_coding}, and served as a proof of concept.
Several follow-up works~\cite{wavenet_vqvae_coding, samplernn_coding} improved quality and computational complexity, and~\cite{lpcnet_coding} presented \ac{lpcnet}, a low complexity decoder which synthesizes good quality clean speech at \kbps{1.6}.
We have shown in our previous work~\cite{ssmgan} that the same bitstream used in \ac{lpcnet} can be decoded by \ac{ssmgan}, a feed-forward GAN model, which provides significantly better quality.

All of these models produce high-quality clean speech, but are not robust in the presence of noise and reverberation. 
\ac{lyra}~\cite{lyra} was the first model to directly address this problem.
Overall, it seems that the generalization capabilities and the quality of level 2 models are partly weakened by the limitations of the classical representation of speech at the encoder side. 

Many approaches tackling the problem from the perspective of a level 3 solution were proposed~\cite{firstnn_coding, knet_coding, cascade_coding, collabquant_coding}, but these models usually do not target very low bit rates.

\ac{sstream}~\cite{sstream} was the first fully end-to-end approach, operating at low bit rates and robust under many different noise conditions.
It is built on a U-Net convolutional encoder-decoder, without skip connections, and using a residual quantization in the bottleneck.
According to the authors' evaluation \ac{sstream} is stable under a wide range of real-life coding scenarios.
Moreover, it permits to synthesize speech at bit rates ranging from \kbps{3} to \kbps{12}.
Finally, \ac{sstream} works at \khz{24}, implements a noise reduction mode, and can also code music.
More recently the work~\cite{tfnet} presented another level 3 solution using a different set of techniques.

We present \ac{ourmodel}, a new model capable of robustly coding wideband speech at \kbps{3}.
The architecture behind \ac{ourmodel} is fundamentally different from \ac{sstream} and is the main aspect of novelty of our approach.
The encoder architecture is based on our proposed \ac{frontend}, which uses a sandwich of convolutional and recurrent layers to efficiently model intra-frame and inter-frame dependencies.
The \ac{frontend} layer is followed by a series of convolutional residual blocks with no downsampling and by a residual quantization.
The decoder architecture is composed of a recurrent neural network followed by the decoder of \ac{ssmgan}.

We show that data augmentation can significantly improve robustness against a wide range of different types of noises and reverberation.
We extensively test our model with many types of signal perturbations and unseen speakers as well as unseen languages.
Moreover, we analyze the unsupervised speech signal clustering achieved by the latent representation.
Our contributions are the following:
\begin{itemize}
\item We introduce \ac{ourmodel}, a new end-to-end neural codec for speech.
\item We present the \ac{frontend} layer, which offers an efficient way of exploiting intra and inter-frame dependencies, for learning a latent representation suitable for quantization.
\item We analyze some interesting clustering behaviour exhibited by \ac{ourmodel}'s quantized latent.
\item We demonstrate \ac{ourmodel}'s robustness against many types of noise and reverberation, via objective and subjective evaluations.
\end{itemize}

\section{Proposed Architecture}
\label{sec:neural_spco}

\begin{figure}[t]
  \centering
  \includegraphics[width=\linewidth]{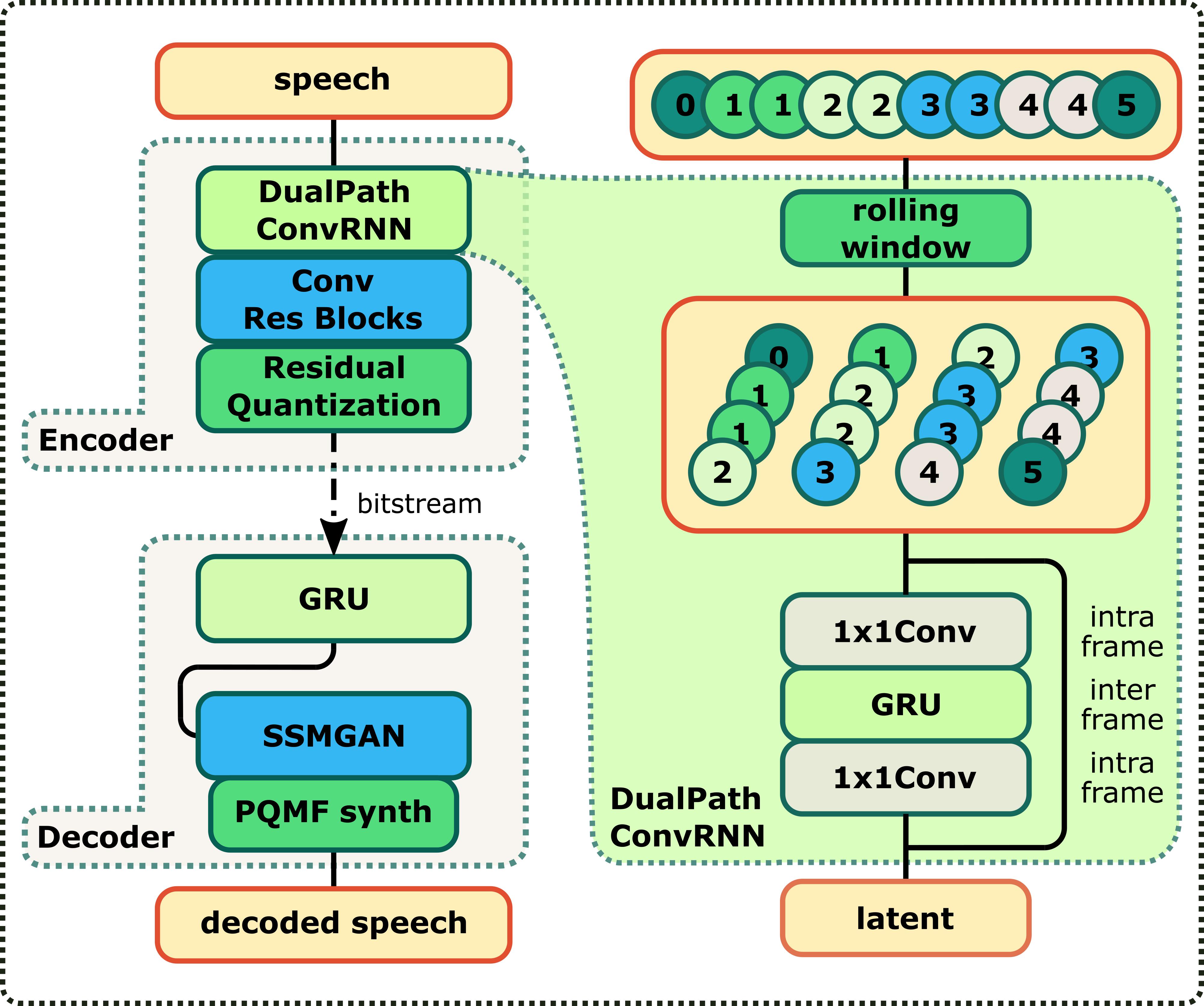}
  \caption{\ac{ourmodel}'s high level architecture and the \ac{frontend}.}
  \label{fig:complete_model}
\end{figure}

As illustrated in Fig.~\ref{fig:complete_model}, the proposed model consists of a learned encoder, a learned quantization layer and a recurrent pre-net followed by a \ac{ssmgan} decoder.

The encoder architecture counts $2.09$ M parameters, whereas the decoder has $3.93$ M parameters.
The encoder rarely reuses the same parameters in computation, as we hypothesize that this favors generalization.
It runs around 40x faster than real time on a single thread of an Intel(R) Core(TM) i7-6700 CPU at 3.40GHz.
The decoder runs around 2x faster than real time on the same architecture, despite only having double as many parameters as the encoder.
Our implementations and design are not optimized for inference speed.

\subsection{Encoder}
\label{subsec:encoder}

The encoder architecture relies on our newly proposed \ac{frontend}, which was inspired by~\cite{dualpathrnn}.
This layer consists of a rolling window operation followed by a 1x1-convolution with $512$ channels, a GRU with $128$ hidden dimensions, and finally another 1x1-convolution with $256$ channels, all activated via LeakyReLUs.
The rolling window transform reshapes the input signal of shape $[1, t]$ into a signal of shape $[s, f]$, where $s$ is the length of a frame and $f$ is the number of frames.
We use frames of \ms{10} with \ms{5} from the past frame and \ms{5} lookahead.
For \seconds{2} of audio at \khz{16} this results in $s = 80 + 160 + 80 = 320$ samples and $f = 200$.
The 1x1-convolutional layers model the time dependencies within each frames, i.e. intra-frame dependencies, whereas the GRU models the dependencies between different frames, i.e. inter-frame dependencies.
This approach allows us to avoid downsampling via strided convolutions or interpolation layers, which in early experiments were shown to strongly affect the final quality of the audio synthesized by \ac{ssmgan}.

The rest of the encoder architecture consists of $4$ residual blocks, each of which consists of a 1d-convolution with kernel size $3$ followed by a 1x1-convolution, both with $256$ channels and activated via LeakyReLUs.
The use of the \ac{frontend} provides a compact and efficient way to model the temporal dependencies of the signal, hence making the use of dilation or other tricks for extending the receptive field of the residual blocks unnecessary.

\subsection{Quantization}
\label{subsec:quantization}

The encoder architecture produces a latent vector of dimension $256$ for each packet of \ms{10}.
This vector is then quantized using a learned residual vector quantizer based on \ac{vqvae}~\cite{vqvae} as in \cite{sstream}.
In a nutshell, this quantizer learns multiple codebooks on the vector space of the encoder latent packets.
The first codebook approximates the latent output of the encoder $z = E(x)$ by the closest entry of the codebook $z_e$.
The second codebook does the same on the residual of the quantization, i.e. on $z - z_e$, and so on for the following codebooks.
This technique is well-known in classical coding, and permits to use the vector space structure of the latent to code many more points with significantly less complexity than by using a single codebook of equivalent bit rate.

In \ac{ourmodel} we use a residual quantizer with three codebooks each at $\SI{10}{bits}$ to code a packet of \ms{10}, hence resulting in a total of \kbps{3}.
It is important to note that during inference, it is possible to drop one or two codebook indices and get an approximation of the final output.
\ac{ourmodel} delivers then a scalable bitstream from \kbps{1} to \kbps{3}.
Since we did not explicitly train it for the intermediate bit rates, we did not include them in our final subjective evaluation, even though the signal is coherent and intelligible at each bit rate.

\subsection{Decoder}
\label{subsec:decoder}

The decoder architecture is composed of a recurrent neural network followed by a \ac{ssmgan} decoder.
We use a single causal GRU layer as a pre-net in order to prepare the bitstream before feeding it to the \ac{ssmgan} decoder.
We do not apply significant modifications to the \ac{ssmgan} decoder, except for the use of a constant prior signal and the use of $256$ conditioning channels provided by the output of the GRU.
We refer to~\cite{ssmgan} for more details on this architecture.
Briefly, this is a convolutional decoder, which is based on Temporal Adaptive DE-normalization layers (TADE), similar to the FiLM layers used in~\cite{sstream}.
It upsamples the bitstream with very low upsampling scales, and provides the conditioning information at each layer of upsampling, while shaping the signal via softmax-gated tanh activations.

\ac{ssmgan} outputs four Pseudo Quadrature Mirror Filter-bank (PQMF)~\cite{pqmf} sub-bands, which are then synthesized using a synthesis filter.
This filter has 50 samples of lookahead, effectively introducing one frame of delay in our implementation.
The total delay of our system is then \ms{25}, \ms{15} from the encoder and the framing and \ms{10} from the decoder.

\section{Evaluation}
\label{sec:evaluation}

\subsection{Experimental setup}
\label{subsec:experimental_setup}
We train \ac{ourmodel} on the $260$ hours of speech from the LibriTTS Dataset~\cite{libritts} at \khz{16}.
The clean speech samples are augmented by adding background noise from the DNS Challenge Dataset~\cite{dnsnoise} at a random SNR between \db{0} and \db{50}, and also adding reverberation by convolving real or generated room impulse responses (RIRs) from the SLR28 Dataset~\cite{openslr, opensrl_paper}.

The training of \ac{ourmodel} is very similar to the training of \ac{ssmgan} as described in~\cite{ssmgan}.
We first pre-train encoder and decoder together for around $500$k iterations having the spectral reconstruction loss of~\cite{pwgan} and the MSE loss as objective.
We then turn on the adversarial loss and the discriminator feature losses from~\cite{melgan} and train for another $700$k iterations; beyond that we have not seen substantial improvements.
The generator is trained on audio segments of \seconds{2} with batch size $64$.
We use an Adam~\cite{adam} optimizer with learning rate $1\cdot 10^{-4}$ for the pre-training of the generator, and decrease the learning rate to $5\cdot 10^{-5}$ as soon as the adversarial training starts.
We use an Adam optimizer with learning rate $2\cdot 10^{-4}$ for the discriminator.

\subsection{Qualitative statistical analysis of the latent}
\label{subsec:tsne}

\begin{figure}[t]
  \centering
  \includegraphics[width=\linewidth]{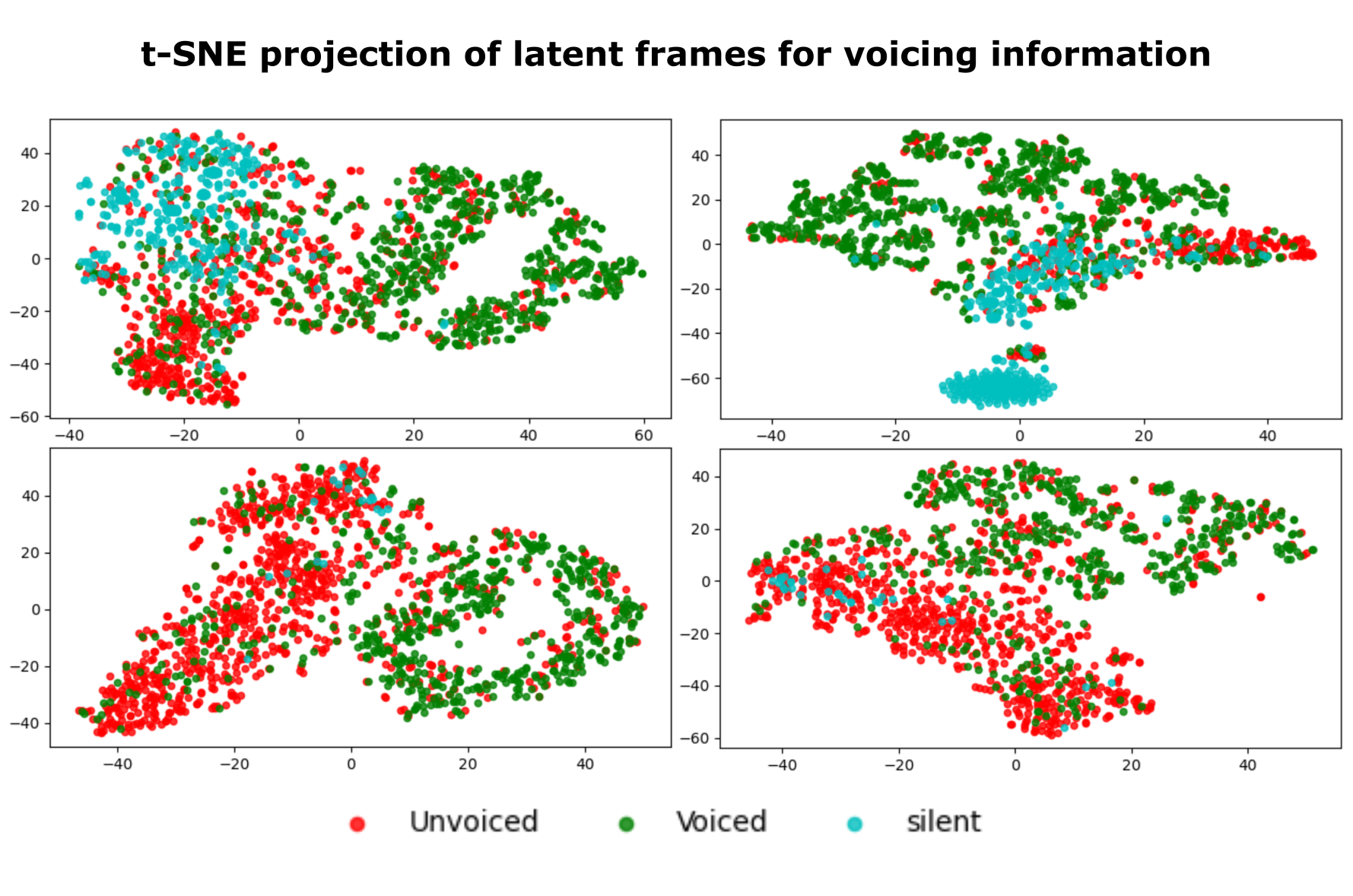}
  \caption{Voiced and unvoiced frames are clustered. Each subplot represents a different speaker selected at random.}
  \label{fig:tsne_voiced}
\end{figure}

\begin{figure}[t]
  \centering
  \includegraphics[width=\linewidth]{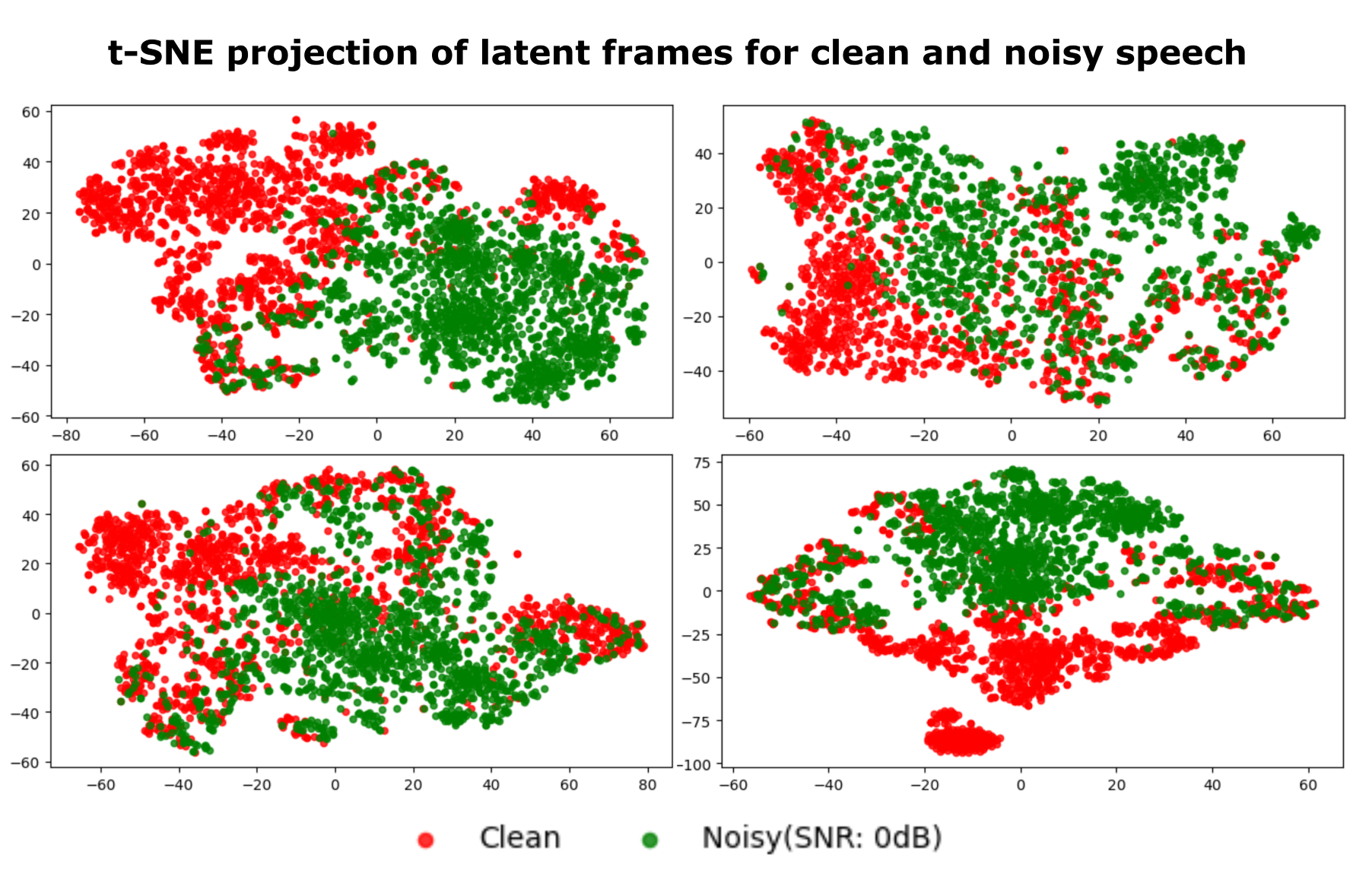}
  \caption{Noisy and clean frames are clustered. Each subplot represents a different speaker selected at random.}
  \label{fig:tsne_noise}
\end{figure}

\begin{figure}[t]
  \centering
  \includegraphics[width=\linewidth]{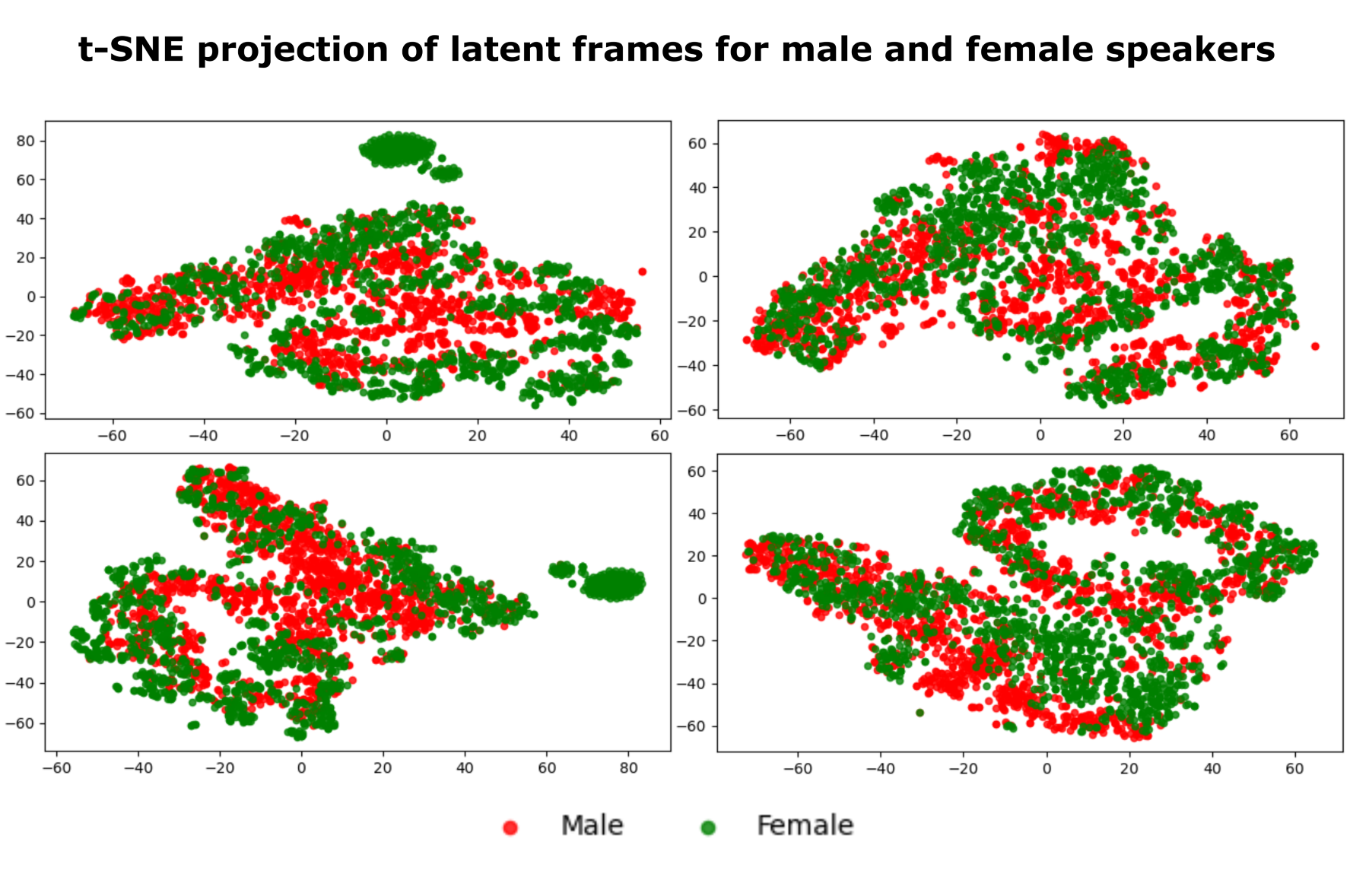}
  \caption{No particular clustering is shown. Each subplot represents a different speaker combination selected at random.}
  \label{fig:tsne_gender}
\end{figure}

We provide a qualitative analysis of the distribution of the latent in order to give a better understanding of its behaviour in practice.
The quantized latent frames are embedded in a space of dimension $256$, hence in order to plot their distribution we use their t-SNE projections~\cite{tsne}.
For each experiment, we first encode the audio with different recording conditions and we label each frame depending on a priori information regarding its acoustic or linguistic characteristics; finally we look for clusters in the low dimensional projections.
Each subplot represents audio coming from different speakers randomly selected from the LibriTTS Dataset.
Notice that the model is not trained with any clustering objective, hence any such behaviour shown at inference time is an emergent aspect of the training set up.

In our first experiment we test voicing information using a VAD algorithm to label each frame automatically.
Fig.~\ref{fig:tsne_voiced} shows a clear clustering of voiced, unvoiced, and silent frames.
We also analyze the clustering ability of the pitch level, but we do not observe a clear trend and therefore we do not add the plots due to lack of space.

In Fig.~\ref{fig:tsne_noise} noise injection during data augmentation is analyzed.
We once again notice a clear division between noise frames and clean frames in the latent space, suggesting that the model is using distinct parts of the latent for these distinct modes.

Finally, we test linguistic and speaker dependent characteristics such as gender (Fig.~\ref{fig:tsne_gender}), speaker id, and language (figures not shown due to lack of space).
In these cases we do not observe any particular clusterings, suggesting that the model is not able to distinguish between these macro-level aspects.

We hypothesize that these clustering behaviours might reflect the compression strategy of the model, which would be in line with well-known heuristics already in use in classical codecs.

\subsection{Objective scores}
\label{subsec:obj_scores}

We evaluate \ac{ourmodel} using several objective metrics.
It is well-known that such metrics are not reliable for assessing the quality of neural codecs~\cite{wavenet_coding, ssmgan}, as they disproportionately favor waveform-preserving codecs.
Nonetheless, we report their values for comparison purposes.
We consider ViSQOL v3~\cite{visqol}, POLQA~\cite{polqa} and the speech intelligibility measure STOI~\cite{stoi}.

The scores are calculated on two internally curated test sets, the StudioSet and the InformalSet, respectively in Table~\ref{tab:studio_set} and \ref{tab:informal_set}.
The StudioSet is constituted of 108 multi-lingual samples from the NTT Multi-Lingual Speech Database for Telephonometry, totalling around 14 minutes of studio-quality recordings.
The InformalSet is constituted of 140 multi-lingual samples scraped from several datasets including LibriVox, and totalling around 14 minutes of audio recordings.
This test set includes samples recorded with integrated microphones, more spontaneous speech, sometimes with low background noise or reverberation from a small room.
\ac{ourmodel} scores the best among the neural coding solutions across all three metrics.

\begin{table}[th]
  \caption{Average objective scores for neural decoders on the StudioSet. For all metrics higher scores are better. Confidence intervals are negligible for POLQA and ViSQOL v3, while for STOI they are smaller than $0.02$.}
  \label{tab:studio_set}
  \centering
  \begin{tabular}{lccc}
     \toprule
     Codec & POLQA & STOI & ViSQOL v3 \\
     \midrule
     \ac{opus} \kbps{6} & 1.681 & 0.480 & 2.273 \\
     \ac{evs} \kbps{5.9} & \textbf{3.308} & \textbf{0.553} & \textbf{3.036} \\
     \midrule
     \ac{ssmgan} \kbps{1.6} & 2.213 & 0.536 & 2.505 \\
     \ac{ourmodel} \kbps{1} & 1.534 & 0.612 & 2.109 \\
     \ac{ourmodel} \kbps{2} & 2.382 & 0.641 & 2.615 \\
     \ac{ourmodel} \kbps{3} & \textbf{2.548} & \textbf{0.643} & \textbf{2.841} \\
     \bottomrule
  \end{tabular}
\end{table}

\begin{table}[th]
  \caption{Average objective scores for neural decoders on the InformalSet. For all metrics higher scores are better. Confidence intervals are negligible for POLQA and ViSQOL v3, while for STOI they are smaller than $0.025$.}
  \label{tab:informal_set}
  \centering
  \begin{tabular}{lccc}
     \toprule
     Codec & POLQA & STOI & ViSQOL v3 \\
     \midrule
     \ac{opus} \kbps{6} & 1.833 & 0.613  & 2.357 \\
     \ac{evs} \kbps{5.9} & \textbf{3.486} & \textbf{0.736} & \textbf{3.071} \\
     \midrule
     \ac{ssmgan} \kbps{1.6} & 2.267 & 0.647 & 2.476 \\
     \ac{ourmodel} \kbps{1} & 1.659 & 0.745 & 2.074 \\
     \ac{ourmodel} \kbps{2} & 2.492 & 0.802 & 2.595 \\
     \ac{ourmodel} \kbps{3} & \textbf{2.707} & \textbf{0.817} & \textbf{2.822} \\
     \bottomrule
  \end{tabular}
\end{table}

\subsection{Subjective Evaluation}
\label{subsec:lt}

We subjectively test the model on challenging unseen conditions in order to assess its robustness.
We select a test set of speech samples from the NTT Dataset comprising unseen speakers, languages and recording conditions.
In the naming convention "m" stands for male, "f" for female, "ar" for Arabic, "en" for English, "fr" for French, "ge" for German, "ko" for Korean, and "th" for Thai.

We also test the model on noisy speech, for this we select the same speech samples as for the clean speech test, and apply a similar augmentation policy as in Section~\ref{subsec:experimental_setup}.
We mix ambient noises (e.g. airport noises, typing noises, ...) at SNRs between \db{10} and \db{30}, and then convolve with room impulse responses (RIR) coming from small, medium and large rectangular rooms.
More precisely, "ar/f", "en/f", "fr/m", "ko/m", and "th/f" are convolved with RIRs from small rooms, and hence for these signals the reverberation does not play a big role; whereas the other samples are convolved with RIRs medium and large size rooms.
Notice that the SNR interval is different from the one used in training as SNRs lower than \db{10} would have been too challenging to evaluate, and SNRs above \db{30} would have not provided enough perturbation in the signal.

\begin{figure}[t]
  \centering
  \includegraphics[width=\linewidth]{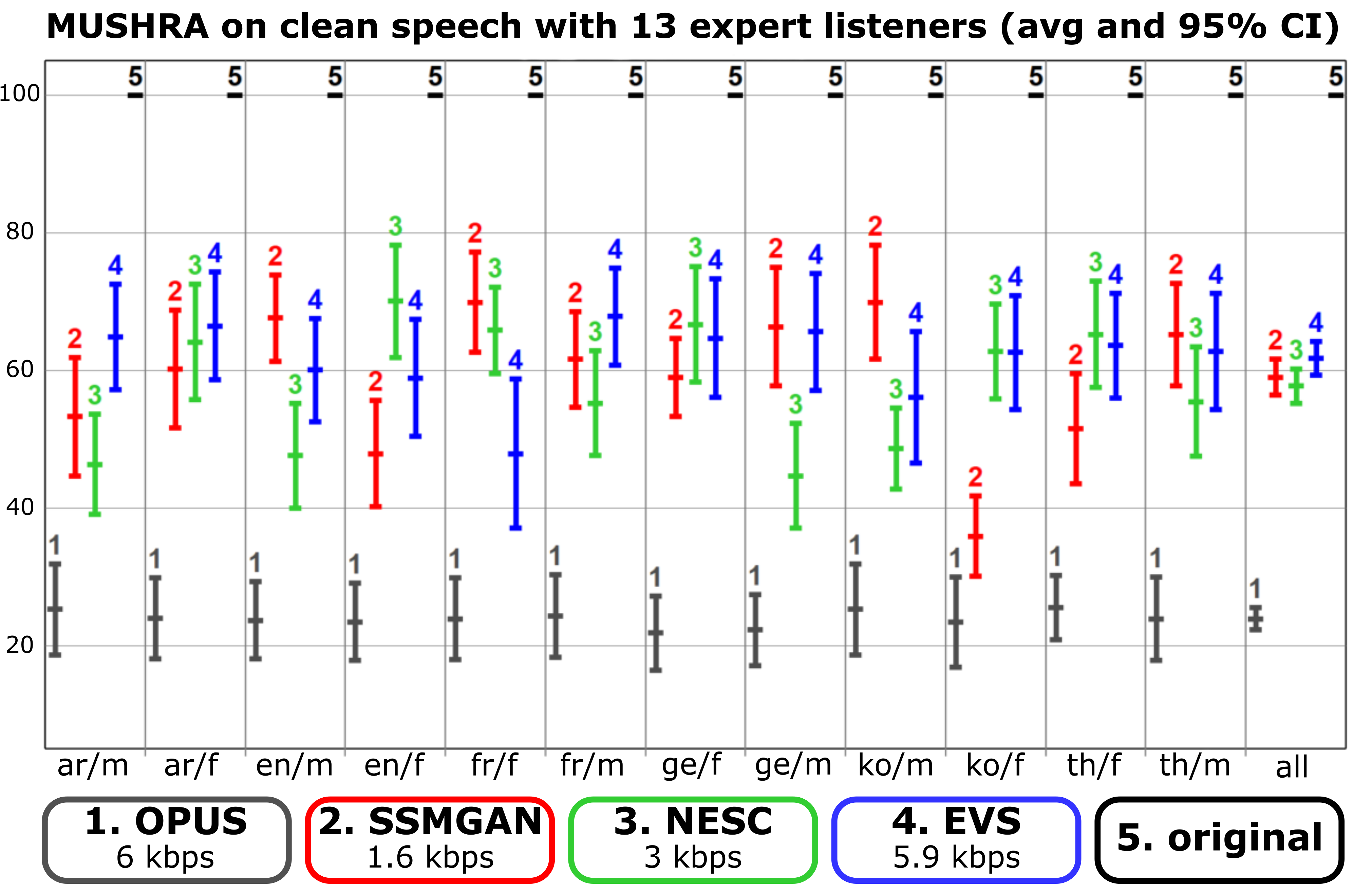}
  \caption{The listening test on clean speech shows that \ac{ourmodel} is on par with \ac{evs} and \ac{ssmgan}.}
  \label{fig:lt_clean}
\end{figure}

\begin{figure}[t]
  \centering
  \includegraphics[width=\linewidth]{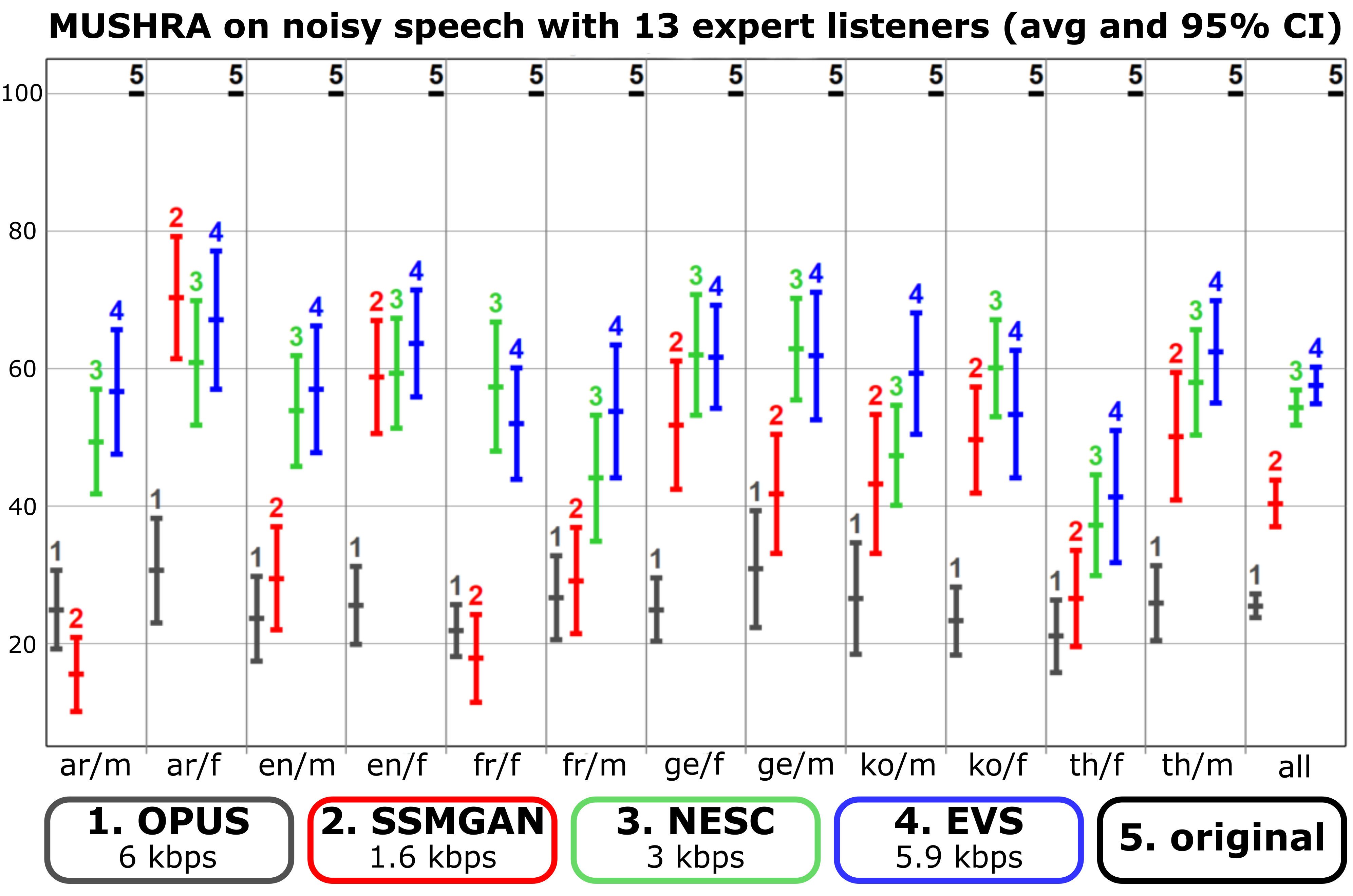}
  \caption{The listening test on noisy speech shows that \ac{ourmodel} is robust under very challenging conditions.}
  \label{fig:lt_noisy}
\end{figure}

We conduct two MUSHRA~\cite{mushra} listening tests, both involving 13 expert listeners.
The test results for clean speech in Fig.~\ref{fig:lt_clean} show that \ac{ourmodel} is on par with \ac{ssmgan} and \ac{evs}~\cite{evs} in this case.
The test results for noisy speech in Fig.~\ref{fig:lt_noisy} confirm that \ac{ssmgan} is not robust to such scenarios, while \ac{ourmodel} competes with \ac{evs} in this case.

The anchor for the tests is generated using \ac{opus}~\cite{opus} at \kbps{6}, since the quality is expected to be very low at this bit rate.
We took \ac{evs}~\cite{evs} at \kbps{5.9} nominal bit rate as good quality benchmark for the classical codecs.
In order to avoid an influence of CNG frames with different signature on the test, we deactivated the DTX transmission.
Finally, we test our solution against our previous neural decoder \ac{ssmgan} at \kbps{1.6}.
\ac{ssmgan} at \kbps{1.6} is trained on the VCTK Corpus~\cite{vctk}, hence the comparison with \ac{ourmodel} is not completely fair.
Early experiments showed that training \ac{ssmgan} at \kbps{1.6} with noisy data is more challenging than expected.
We suppose that this issue is due to its reliance on the pitch information, which might be challenging to estimate in noisy environments.
For this reason we decided to test \ac{ourmodel} against the best neural clean speech decoder that we have access to, namely \ac{ssmgan} trained on VCTK, and still add it to the noisy speech test as an additional condition to show its limitations.

Both tests show that \ac{ourmodel} is on par with \ac{evs}, while having half of its bit rate.
Fig.~\ref{fig:lt_noisy} shows the limitations of \ac{ssmgan} when working with noisy signals and confirms that the quality of \ac{ourmodel} stays high even in these challenging conditions\footnote{Check our demo samples at: https://fhgspco.github.io/nesc/}.

\section{Conclusions}
\label{sec:conclusion}

We present \ac{ourmodel}, a new GAN model capable of high-quality and robust end-to-end speech coding.
It relies on the new \ac{frontend} as the main building block for efficient and reliable encoding.
We test our setup via objective quality measures and subjective listening tests, and show that \ac{ourmodel} is robust under various types of noise and reverberation.
We show a qualitative analysis of the latent structure giving a glimpse of the internal workings of our codec.
Future work will be directed toward further complexity reduction and quality improvements.

\bibliographystyle{IEEEtran}

\bibliography{mybib}

\end{document}